# Pneumonia Diagnosis through pixels - A Deep Learning Model for detection and classification

Gurpur Amit Karanth
School of Electronics and Communication
Vellore Institute of Technology
Chennai, India

Janani S
School of Electronics and Communication
Vellore Institute of Technology
Chennai, India

Ajeetha B
School of Electronics and Communication
Vellore Institute of Technology
Chennai, India

Dr.Brintha Therese A
Professor
Vellore Institute of Technology
Chennai, India

Dr. Rajeswaran Rangasami
Professor
Sri Ramachandra University
Chennai, India

*Abstract - Manual identification and classification of pneumonia and COVID-19 infection is a cumbersome process that, if delayed can cause irreversible damage to the patient. We have compiled CT scan images from various sources, namely, from the China Consortium of Chest CT Image Investigation (CC-CCII), the Negin Radiology located at Sari in Iran, an open access COVID-19 repository from Havard dataverse, and Sri Ramachandra University, Chennai, India. The images were preprocessed using various methods such as normalization, sharpening, median filter application, binarizing, and cropping to ensure uniformity while training the models. We present an ensemble classification approach using deep learning and machine learning methods to classify patients with the said diseases. Our ensemble model uses pre-trained networks such as ResNet-18 and ResNet-50 for classification and MobileNetV2 for feature extraction. The features from MobileNetV2 are used by the gradient-boosting classifier for the classification of patients. Using ResNet-18, ResNet-50, and the MobileNetV2 aided gradient boosting classifier, we propose an ensemble model with an accuracy of 98 percent on unseen data.*

*Keywords: Pneumonia, Lung CT, Deep Learning, Convolutional Neural Networks, Medical Imaging, Computer-Aided Diagnosis.*

## I. INTRODUCTION

Pneumonia, a leading cause of death worldwide, is a medical condition that inflames the air sacs in one or both lungs. Pneumonia can affect people of all ages and range in severity from moderate to life-threatening. It continues to be a major global health concern, causing an estimated 2.5 million fatalities each year, with a disproportionately high impact on young children and the elderly. Improved patient outcomes and efficacious therapy depend on an early and precise diagnosis. Chest X-rays have historically been used to diagnose pneumonia in its early stages. They may, however, be insensitive and specific, which could result in a false diagnosis. This is when lung CT (computed tomography) scans come in handy. Lung CT scans offer finely detailed cross-sectional images of the lungs, making abnormalities related to pneumonia easier to see.

A model that can automatically identify pneumonia in new CT scans by training a CNN (Convolutional Neural Networks) on a sizable dataset of lung CT scans with confirmed pneumonia and healthy lung visuals is created. This approach makes use of ensemble learning, a potent machine learning methodology that enhances performance by combining the advantages of several models.

The three different classification algorithms that were used to classify pneumonia, COVID infection, and normal patients are:
- Pre-trained CNN models:
    - ResNet-18
    - ResNet-50
    - MobileNetV2 (feature extraction)
- Gradient Boosting Classifier

## II. Literature review

An effective ensemble-based CAD system for pneumonia detection in chest X-ray images, showcasing high accuracy but recognizing the need for potential improvements in occasional mispredictions and computational efficiency is introduced. Their research showed encouraging outcomes. However, more study is still required to test and modify these techniques for lung CT pictures in order to improve diagnostic capabilities, considering the various modalities and complexity of lung CT scans [1].

Deep learning models, particularly Convolutional Neural Networks (CNNs), demonstrate superior performance in automating the diagnosis process compared to traditional machine learning approaches. In order to improve efficiency and accuracy, transfer learning techniques are used to leverage pre-trained models and modify them for COVID-19 and pneumonia detection tasks. By using the F-RNN-LSTM system, a better accuracy of 95% with low computational efforts is achieved [2].

A novel method for textural feature-based automatic pneumonia diagnosis in chest X-ray images is explored. The images' textural characteristics were extracted in order to identify minute patterns that point to the existence of pneumonia. This method provides a non-invasive and effective method to diagnose pneumonia, important for prompt medical attention. Compared to conventional techniques, the use of textural cues improves the sensitivity and specificity of pneumonia identification. By using three machine learning algorithms: KNN, SVM, and RF, improvements are shown in both accuracy and F-Score between 4% and 8% [3].

Using datasets from CT scans and X-ray images, the research suggests a unique ensemble framework that combines deep learning models for better COVID-19 and pneumonia identification. The study highlights the ensemble framework's high diagnostic precision in identifying COVID-19 cases, pneumonia, and healthy individuals through thorough analysis, enabling timely and precise patient management. Ensemble (VGG-16, ResNet, DenseNet) attains 95-96% accuracy in COVID-19 and pneumonia classification in chest imaging, outperforming with under 0.5s time complexity [4].

A deep learning framework for interpreting chest X-ray pictures that use several CNN

architectures, such as AlexNet, and VGG16, among others is studied. It emphasizes the importance of transfer learning, where pre-trained models are fine-tuned for the specific task of pneumonia detection. Modified deep convolutional neural network (DCNN) architecture is used with a training accuracy of 98.02%, and validation accuracy of 96.09%, which is much higher than the existing approaches and techniques [5].

Chest X-ray pictures are used to show how well CNNs distinguish COVID-19 pneumonia from other kinds of pneumonia. By using pre-trained CNN architectures like DenseNet-201 in conjunction with transfer learning, they were able to classify COVID-19 pneumonia cases with high accuracy. DenseNet-201 CNN produced about 94.96% accuracy in COVID-19 screening from CXIs which promises faster diagnosis, lower radiation, and cost-effective healthcare[6].

AI systems show a great deal of promise for helping to quickly and accurately diagnose COVID-19 pneumonia from chest X-ray pictures. The high sensitivity and specificity of these AI-based systems enable early detection and treatment initiation, which is essential for restricting the disease's spread. The AI models for COVID-19 pneumonia detection using CXR images, show high sensitivity and scalability for remote screening, providing practical solutions for pulmonary diseases [7].

### III. METHODOLOGY

This section outlines the methodology of our research paper investigating pneumonia diagnosis with lung CT scans using an ensemble model incorporating the pre-trained CNNs - ResNet-18 and ResNet-50, and the Gradient-boosting classifier. The block diagram in Fig 3.1 aims to explain the flow of the model proposed.

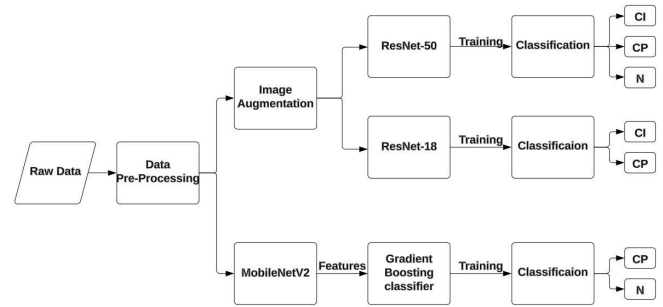

Fig 3.1 Block diagram of the proposed model

This model uses lung CT scans of healthy and infected patients to explore deep learning for pneumonia diagnosis. We obtained lung CT scans from the China Consortium of Chest CT Image Investigation (CC-CCII), Negin Radiology located at Sari in Iran, the Harvard Dataverse, a repository with confirmed COVID-19 cases, and from Sri Ramachandra University, Chennai, India.

Next, in order to ensure consistency, we preprocess the data using techniques such as image enhancement, selection, noise removal using median filtering, and cropping. The features are extracted using the convolutional layers present in the deep learning models. We have also introduced data augmentation to the dataset to bring in variability to the dataset. The variability introduced will help negate any effects of having a small and constant dataset.

The pre-trained models used in this project are the ResNet-50, ResNet-18, and MobileNetv2. The ResNet-50 and ResNet-18 networks have been trained to classify the input images into different classes, which will be discussed later in the paper. For feature extraction for the Gradient-boosting classifier, we have used pre-trained networks such as MobileNetV2.

Lung segmentation-based characteristics are used by ensemble classifier-Gradient Boosting. After feature extraction, we develop and train the classifier system. Convolutional and pooling layers are used by the CNN to extract spatial characteristics from images, and ensemble classifiers combine several decision trees to increase accuracy. Lastly, we assess each model's performance on a different test dataset using metrics like accuracy, precision, recall, and F1-score.

## IV. PRE-PROCESSING

The raw lung CT scans appear dark and lack visual details due to variations in tissue density. Because of differences in tissue density, raw CT scans of the lungs seem black and lack visual features. Preprocessing techniques are essential in order to improve the image and highlight anatomical features of the lungs for additional analysis.

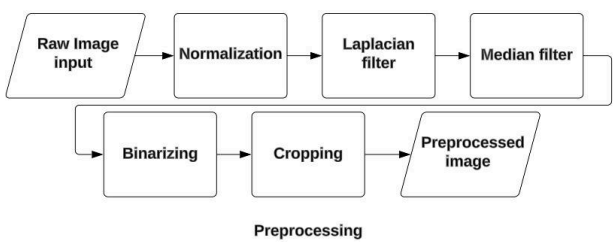

Fig 4.1 Block diagram of pre-processing

A. Image Enhancement

The images were obtained from different sources and were present in different formats. The raw images from Negin radiology are in 16-bit uint grayscale, TIFF format, making them appear pitch black on normal monitors, whereas the images from the Harvard Database and SRMC, Chennai were in DICOM format, where an extra step of converting these images to TIFF was required before normalizing them. The enhancer helps make these images visible on regular monitors as shown in Fig 4.2.1. Each pixel value is normalized by dividing by the maximum pixel value of the image, thereby the output images have a 32-bit double that can be visualized by regular monitors. These images can be later converted to other formats for better compatibility. We stored these images as uint8 TIFF files.

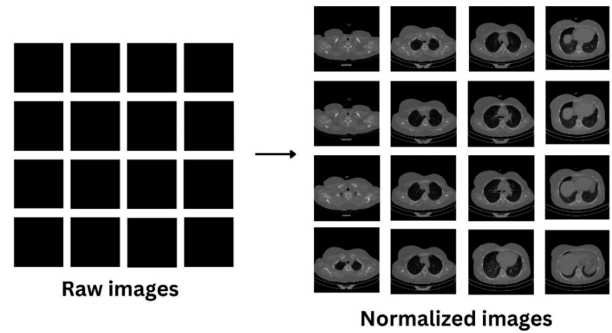

Fig 4.2.1 Normalization of Raw Images

B. Selection

A certain region of interest (ROI) is chosen in every image, in our case we chose 241 to 340 and columns 121 to 370 in the 512x512 image. The selected area in the image contains the lung. Using this ROI, the number of pixels within this region that have an intensity less than a threshold of 200 is chosen. Images that pass this criterion are chosen for further preprocessing, as illustrated in Fig 4.2.2, and the omitted images are discarded as they contain images of the closed lung, which are not useful for classification.

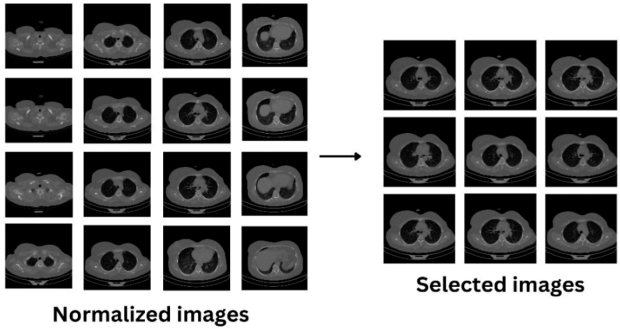

Fig 4.2.2 Selection of Normalized Images

C. Laplacian

Laplacian filters excel at detecting these edges by highlighting areas with rapid intensity changes. This filtering method helps to make regions of infection and minute anomalies more visible in the image. The Laplacian filter is defined as the sum of the second partial derivatives of the image intensity function f(x, y) concerning the spatial coordinates x and y:

$$\nabla^2 f(x, y) = \partial^2 f(x, y) / \partial x^2 + \partial^2 f(x, y) / \partial y^2$$

The filtered image improves structural details and focuses on the edges and contours of the lung, making the images more qualified for training the model(s), as illustrated in Fig 4.2.3

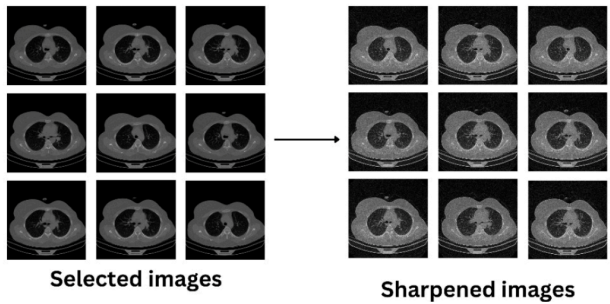

Fig 4.2.3 Sharpening of Selected Images

D. Median Filtering and Binarizing

Reducing noise in lung CT images without substantially blurring or distorting vital elements is the main objective of median filtering. As seen in Fig 4.2.4, the contours are not more pronounced, and the noise has been reduced. Median filtering involves substituting the intensity value of each pixel with the median value of intensity within a local neighborhood determined by a kernel size. This nonlinear procedure works well at reducing noise while preserving edges and minute details.

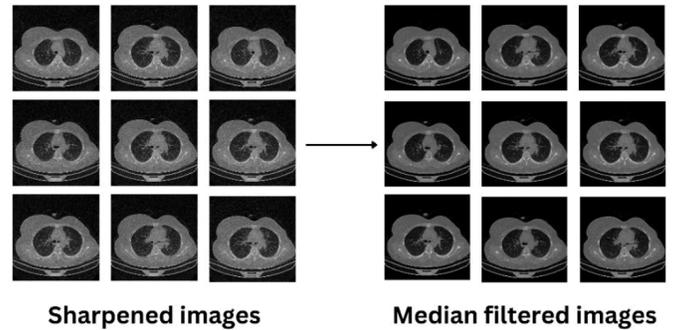

Fig 4.2.4 Median Filtering of Sharpened Images

The median filtered images are binarized to capture minute details and contours of the lung as shown in Fig 4.2.5.

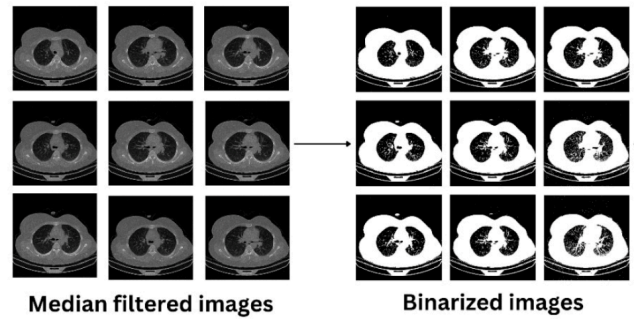

Fig 4.2.5 Binarization of Median Filtered Images

E. Automated Cropping

Noise and unwanted parts of an image often reduce the performance of the model by training it on futile features. Eliminating background pixels and other irrelevant parts of the image will help models concentrate on the lung tissue, which cuts down on the training time and processing requirements and, at the same time, increases classification accuracy. The cropped lung ROI has features that are more pertinent to the diagnosis of pneumonia, which aid in improving model performance. The cropped images are as shown in Fig 4.2.6.

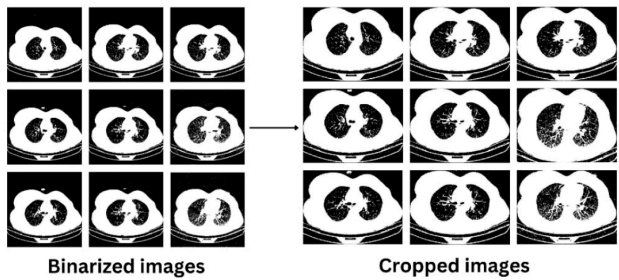

Fig 4.2.6 Cropping of Binarized Images

Cropping lung images with a set dimension will lead to a loss of vital characteristics, leading to poor classification of images. To tackle this, we have automated the cropping process for each image. The input image (1) is first smoothed by using a gaussian filter. On the smoothed image, edge detection is performed by using the sobel operator, as seen in image (2) in Fig. 4.3. The image is dilated first to fill any holes in the lung region, and the remaining gaps are closed as shown in images (3) and (4) of Fig.4.3. Further, the largest connected component is detected to aid in cropping the image. Using the bounding box function in Matlab, the image is cropped and resized to 224x224 to keep all the images uniform.

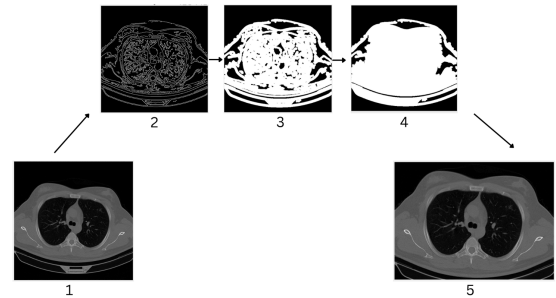

Fig. 4.3 Process of Automated Cropping

**V. PRE TRAINED NETWORKS**

We have utilized pre-trained networks to detect pneumonia rather than building a CNN model from scratch. Using pretrained networks and transfer learning, we can achieve remarkable classification accuracy even with a small amount of labeled data by utilizing the information gathered from the pretrained networks. It improves accuracy, shortens training times, and speeds up convergence in detecting anomalies linked to pneumonia in CT scans by optimizing pre-existing structures. By using well-established feature representations that have been trained from a variety of data sets, this method helps the model generalize better to new datasets and clinical scenarios.

A. Resnet-50 and ResNet-18 - for classification

ResNet-50 and ResNet-18, like other pretrained networks, are trained using over a million images from datasets such as ImageNet. These pretrained models are capable of classifying images into over 1000 classes. Pre-trained networks can be optimized to classify images of classes different from the ones they were trained on. Since both networks have not been trained on medical images, their performance on CT scan images is not optimized for precise classification of medical images. Fig 5.1 and Fig 5.2 show the block diagrams of ResNet-18 and ResNet-50 respectively.

**ResNet-18:** To make the model familiar with medical images and increase its overall capability, the first 44 layers (14 convolution layers, reLu layers etc.) and the last 15 layers (2 convolutional layers) of the model, as in Fig 5.1 are frozen first and trained using 8700 images of patients with COVID-19 infection (49 patients) and of patients with pneumonia (48 patients). After this, the first 59 layers (17 convolutional layers) of the model are frozen, and the last 12 layers (2 convolutional layers and multiple fully connected layers) are kept unfrozen to train the model for a second time. This time, the model is trained using 4600 images of patients with COVID-19 infection (20 patients) and of patients with pneumonia (23 patients). Using this method has made the model familiar with the actual training data, making it more efficient and precise.

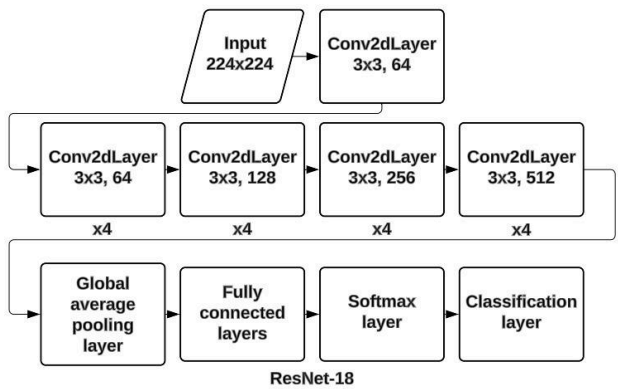

Fig 5.1 Block diagram of ResNet-18

**ResNet-50:** A similar methodology is followed while re-training the ResNet-50. Layers from 1 to 152 are frozen (the first 44 convolutional layers) and the layers from 163 to 183 (the last 2 convolutional layers) are also frozen for future training . The model is trained using 14049 images, with images of patients with COVID-19 infection (49 patients) , pneumonia (48 patients), and normal patients (45 patients). Further, the first 162 layers are frozen, and the layers from 163 to 183 are unfrozen to train a second set of 6826 images, with images of patients with COVID-19 infection (20 patients), pneumonia (23 patients) and normal patients (26 patients).

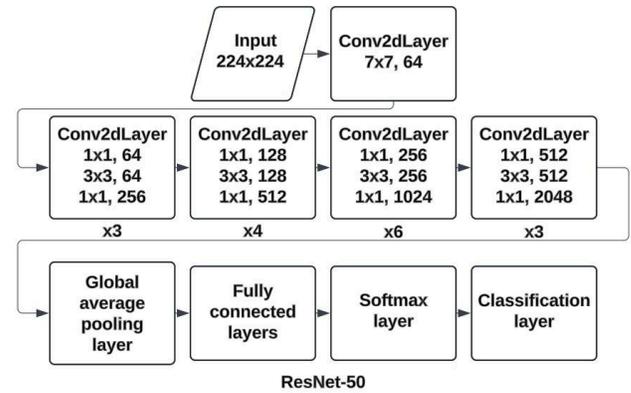

Fig 5.2 Block diagram of ResNet-50

B. MobileNetV2 - for feature extraction

MobileNetV2 is a lightweight architecture, making it computationally efficient for deployment on resource-constrained devices without compromising accuracy. MobileNetV2 is used to extract features for the gradient boosting classifier. The model is trained using approximately 15000 images. In transfer learning with the MobileNetv2 model, as shown in Fig 5.3, it's common practice to freeze all the convolutional layers and only fine-tune the fully connected layers that are added on top of the pre-trained MobileNetv2 model. In MobileNetV2, there are typically 19 blocks, and each block contains one or more convolutional layers. A total of 53 convolutional layers are frozen by setting their weight factor, weight learn rate factor, and bias learn rate factor to zero. Freezing these layers prevents their weights from being updated during training on the new dataset, thus preserving the learned representations. By freezing these layers, we reduce the risk of overfitting on the new dataset and maintain stability.

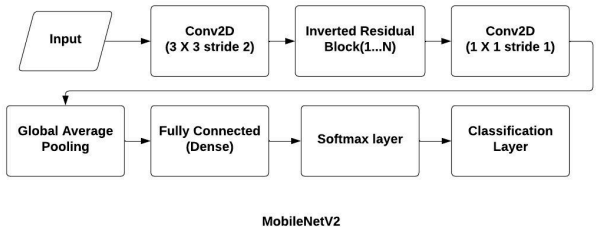

Fig 5.3 Block diagram of MobileNet-V2

## VI. ENSEMBLE MODEL

The idea of building an ensemble model is to leverage the strengths of the individual models. A more reliable and accurate diagnosis will result from combining the predictions of multiple models. This ensemble approach has the potential to be more accurate than relying on a single model.

A. ResNet-50 and ResNet-18

The ResNet-50 model has been trained on all three classes, namely COVID-19 infection, pneumonia, and normal patients. The model is highly accurate in its predictions of COVID-19-infected patients and pneumonia patients and is moderately accurate with its normal patients' predictions. The ResNet-18 has been trained only with two classes: COVID-19 infection and pneumonia, and the model's predictions are high.

B. Gradient Boost Classifier

This method sequentially builds a model by adding decision trees. Each new tree focuses on improving on the previous ones by learning from their errors. It uses the collective predictions of all the trees in the ensemble to determine if the output class label is healthy or pneumonia. The gradient-boosting classifier has been trained on two classes, namely pneumonia and normal patients, and the model achieves an accuracy of 95 percent.

Using the three models, we have developed an ensemble classifier that takes into consideration the predictions of all three classifiers before giving a final result.

## VII. RESULTS AND DISCUSSIONS

A. Training Graph of Pre-trained models

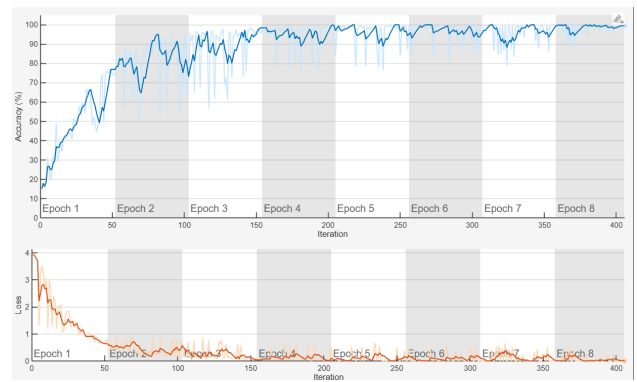

Fig 6.1.1 Training Graph of Resnet-50

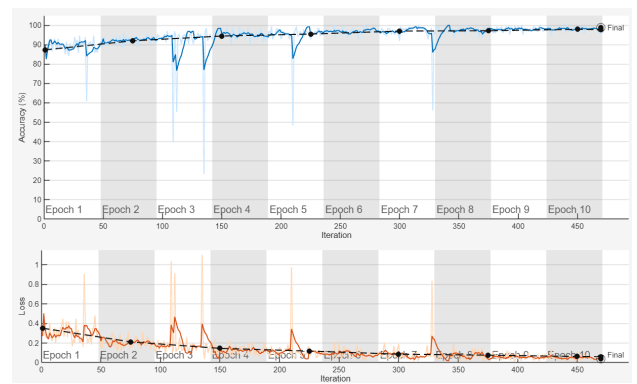

Fig 6.1.2 Training Graph of Resnet-50

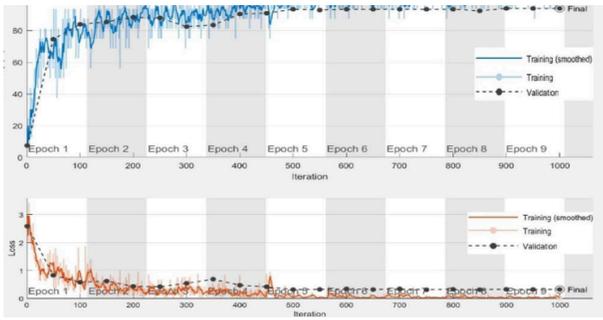

Fig 6.2 Training Graph of MobileNet-V2

From the training graphs in Fig 6.1 and Fig 6.2, we may see that there are some initial fluctuations in loss and accuracy, which denotes that the model is still adjusting its internal parameters. The obtained loss curve is decreasing over epochs and the accuracy curve is increasing over epochs, which is a typical pattern during successful training.

B. ResNet-18 confusion matrix

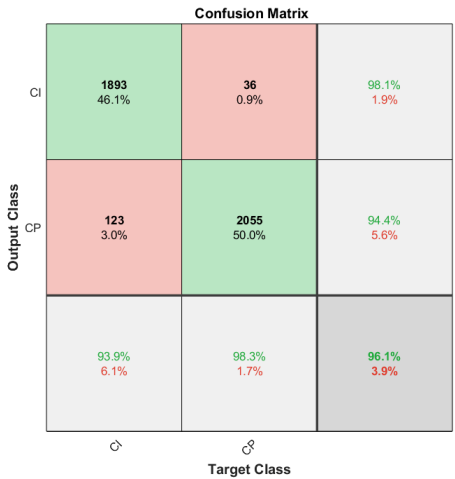

Fig 6.3 Confusion matrix of ResNet-18

From the confusion matrix in Fig 6.3, we derived that the model has an accuracy of 96.1% with a precision of 0.98. The calculated F1 score is 0.96 and the recall is 0.94.

C. ResNet-50 confusion matrix

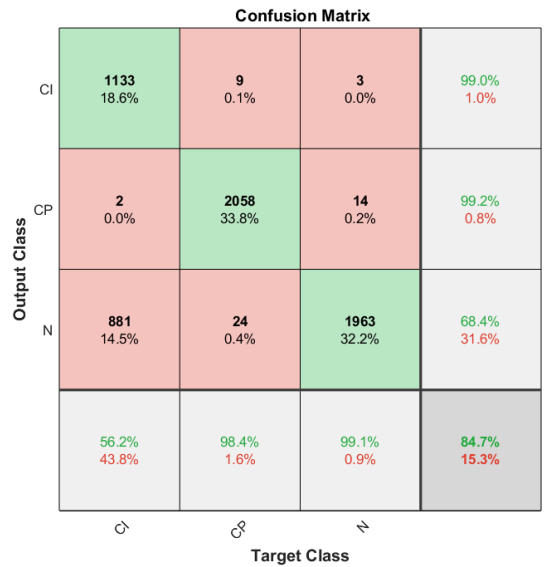

Fig 6.4 Confusion matrix of ResNet-50

From the confusion matrix in Fig 6.4, we derived that the model has an accuracy of 84.7% with a precision rate of 0.98. The calculated F1 score is 0.99 and recall is 0.99.

D. Gradient boosting confusion matrix

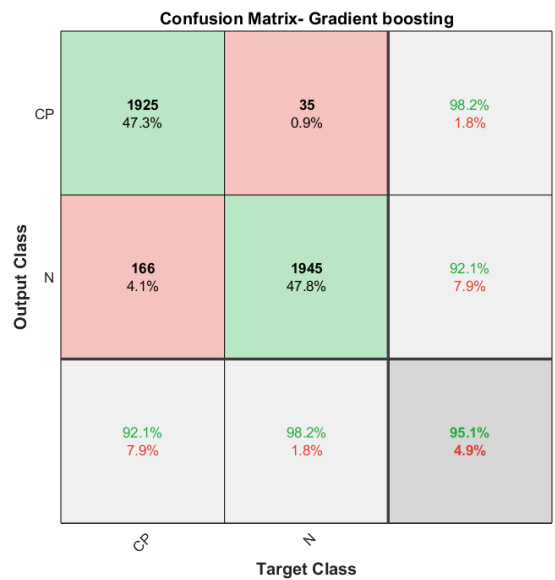

Fig 6.5 Confusion matrix of Gradient Boosting

From the confusion matrix in Fig 6.5, we derived that the model has accuracy of 95.1% with a precision of 0.98. The calculated F1 score is 0.95 and recall is 0.92.

E. Ensemble confusion matrix

Fig 6.6 Confusion matrix of Ensemble Model

From the confusion matrix in Fig 6.6, we derived that the model has accuracy of 98% with a precision rate of 0.97. The calculated F1 score is 0.98 and recall is 0.99.

## VIII. Conclusion

The ensemble model combining pre-trained CNN models and Gradient Boosting Classifier exhibits promising results in pneumonia and COVID-19 infection diagnosis using lung CT scans. When diagnosing pneumonia from lung CT images, the ensemble model that combines CNN and Gradient Boosting Classifier shows promising results.

The proposed model achieved a high accuracy of 98% and a precision of 97%, indicating its strong ability to correctly differentiate between healthy lungs, lungs with pneumonia, and lungs infected with COVID-19. These results suggest that this ensemble approach holds promise as a valuable tool for computer-aided diagnosis in clinical settings. Through the optimization of each algorithm's distinctive features, our model attains improved accuracy, robustness, and versatility. Moreover, the ensemble method reduces the impact of each algorithm's specific flaws, producing predictions that are more reliable.

In medical applications, this ensemble model could assist radiologists in reading CT scans of the lungs, facilitating the early identification and precise diagnosis of pneumonia. Sustained research and development endeavors in this domain have the capability to improve diagnostic procedures, ultimately resulting in enhanced patient outcomes and more effective healthcare distribution.

## REFERENCES

[1] Kundu R, Chakraborty S, Pal A, Sarker A, Shakhawat Hossain SM. Pneumonia detection in chest X-ray images using an ensemble of deep learning models. PLoS One. 2021 Sep 7;16(9):e0256630.

[2] Goyal, S., & Singh, R. (2023). Detection and classification of lung diseases for pneumonia and Covid-19 using machine and deep learning techniques. *Journal of Ambient Intelligence and Humanized Computing*, *14*(4), 3239-3259.

[3] Ortiz-Toro, C., et al. (2022). Automatic detection of pneumonia in chest X-ray images using textural features. *Computers in Biology and Medicine*, *145*, 105466.

[4] Xue, X., et al. (2023). Design and Analysis of a Deep Learning Ensemble Framework Model for the Detection of COVID-19 and Pneumonia Using Large-Scale CT Scan and X-ray Image Datasets. *IEEE Transactions on Medical Imaging*.

[5] Yi, R., et al. (2023). Identification and classification of pneumonia disease using a deep learning-based intelligent computational framework. *Expert Systems with Applications*.

[6] Alhudhaif, A., et al. (2021). Determination of COVID-19 pneumonia based on generalized convolutional neural network model from chest X-ray images. *Computers in Biology and Medicine*.

[7] Baltazar, L. R., et al. (2021). Artificial intelligence on COVID-19 pneumonia detection using chest X-ray images. *PLOS ONE*.

[8] Rahimzadeh, M., Attar, A., & Sakhaei, S. M. (2021). A fully automated deep learning-based network for detecting COVID-19 from a new and large lung CT scan dataset. *Biomedical Signal Processing and Control*, *68*, 102588.

[9] Ahamed, K. U., Islam, M., Uddin, A., Akhter, A., Paul, B. K., Yousuf, M. A., ... & Moni, M. A. (2021). A deep learning approach using effective preprocessing techniques to detect COVID-19 from chest CT-scan and X-ray images. *Computers in biology and medicine*, *139*, 105014.

[10] Kundu, R., Das, R., Geem, Z. W., Han, G. T., & Sarkar, R. (2021). Pneumonia detection in chest X-ray images using an ensemble of deep learning models. *PloS one*, *16*(9), e0256630.

[11] Zhang, K., Liu, X., Shen, J., Li, Z., Sang, Y., Wu, X., ... & Wang, G. (2020). Clinically applicable AI system for accurate diagnosis, quantitative measurements, and prognosis of COVID-19 pneumonia using computed tomography. *Cell*, *181*(6), 1423-1433.

[12] Mostafavi, S. M. COVID19-CT-Dataset: An Open-Access Chest CT Image Repository of 1000+ Patients with Confirmed COVID-19 Diagnosis 2021. *DOI: https://doi.org/10.7910/DVN/6ACUZJ*.